\documentclass[twoside,twocolumn,9pt]{article}
\usepackage{import}
\usepackage{preamble}

\begin{document}

%%% TITLE, AUTHORS AND ABSTRACT %%%
%%% TITLE %%%
\twocolumn[
  \begin{@twocolumnfalse}
\vspace{3cm}
\sffamily
\begin{tabular}{m{4.5cm} p{13.5cm} }

\includegraphics{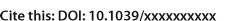} & \noindent\LARGE{\textbf{Laser-induced fragmentation of coronene cations$^\dag$}} \\
\vspace{0.3cm} & \vspace{0.3cm} \\

%%% AUTHORS %%%
 & \noindent\large{Sanjana Panchagnula,$^{1,2,3}$ Jerry Kamer,$^{1}$} Alessandra Candian,$^{4}$ Helgi R. Hrodmarsson,$^{1,5}$ Harold Linnartz,$^{1}$ Jordy Bouwman,$^{1,6,7,8,^{\ast}}$ and Alexander G.G.M. Tielens$^{2,9}$\\

%%% ABSTRACT %%%
\includegraphics{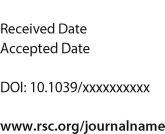} & \noindent\normalsize{Polycyclic aromatic hydrocarbons are an important component of the interstellar medium of galaxies and photochemistry plays a key role in the evolution of these species in space. Here, we explore the photofragmentation behaviour of the coronene cation (\ce{C24H12^.+}) using time-of-flight mass spectrometry. The experiments show photodissociation fragmentation channels including the formation of bare carbon clusters ({\ce{C_n^.+}}) and hydrocarbon chains ({\ce{C_nH_x+}}). The mass spectrum of coronene is dominated by peaks from \ce{C11^.+} and \ce{C7H+}. Density functional theory was used to calculate relative energies, potential dissociation pathways, and possible structures for relevant species. We identify 6-6~\textrightarrow~5-7 ring isomerisation as a key step in the formation of both the bare carbon clusters and the hydrocarbon chains observed in this study. We present the dissociation mechanism outlined here as a potential formation route for \ce{C60} and other astrochemically relevant species.}\\

%Time-of-flight mass spectra of the coronene cation (\ce{C24H12+}) to study the fragmentation patterns of the parent ion in the high-mass region (\textit{m/z}~=~320-130) and low-mass region (\textit{m/z}~=~129-60) are presented. The experiments show photodissociation fragmentation channels including the formation of bare carbon clusters ({\ce{C_n+}}) and hydrocarbon chains ({\ce{C_nH_x+}}). The mass spectrum of coronene is dominated by peaks from \ce{C11+} and \ce{C7H+}. Density functional theory calculations give us relative energies, potential dissociation pathways, and possible structures for relevant species. We identify 6-6~\textrightarrow~5-7 ring isomerisation as a key step in the formation of both, the bare carbon clusters and the hydrocarbon chains observed in this study. We present the bottom-up mechanism proposed here as a potential formation route for \ce{C60} and other astrochemically-relevant species.}\\

\end{tabular}

 \end{@twocolumnfalse} \vspace{0.6cm}]

%%% FONT SETUP - please do not change any commands within this section %%%
\renewcommand*\rmdefault{bch}\normalfont\upshape
\rmfamily
\section*{}
\vspace{-1cm}

%%% FOOTNOTES %%%
\footnotetext{\textbf{1} Laboratory for Astrophysics, Leiden Observatory, Leiden University, 2300 RA, Leiden, The Netherlands.}

\footnotetext{\textbf{2} Leiden Observatory, Leiden University, 2300 RA, Leiden, The Netherlands.}

\footnotetext{\textbf{3} \textit{Current address:} St George's, University of London, Cranmer Terrace, London, SW17 0RE, United Kingdom.}

\footnotetext{\textbf{4} Anton Pannekoek Institute for Astronomy, University of Amsterdam, Science Park 904, 1098 XH, Amsterdam, Netherlands.}

\footnotetext{\textbf{5} \textit{Current address:} Universit{\'e} Paris-Est Cr{\'e}teil and Universit{\'e} Paris Cit{\'e}, CNRS, LISA UMR 7583, F-94010, Cr{\'e}teil, France.}

\footnotetext{\textbf{6} \textit{Current address:} Laboratory for Atmospheric and Space Physics, University of Colorado, CO 80303, Boulder, USA.}

\footnotetext{\textbf{7} \textit{Current address:} Department of Chemistry, University of Colorado, CO 80309, Boulder, USA.}

\footnotetext{\textbf{8} \textit{Current address:} Institute for Modeling Plasma, Atmospheres and Cosmic Dust (IMPACT), NASA/SSERVI, CO 80309, Boulder, USA.}

\footnotetext{\textbf{9} Department of Astronomy, University of Maryland, MD 20742-2421, College Park, USA.}

\footnotetext{\textbf{$^{\ast}$} \textbf{Corresponding author:} \email{jordy.bouwman@colorado.edu}}

\footnotetext{\textbf{\dag} \textbf{Electronic Supplementary Information (ESI) available:} [details of any supplementary information available should be included here]. See DOI: 10.1039/b000000x/}

%%% END OF FOOTNOTES %%%

%%% TRIBUTE %%%
\section*{Tribute to Harold Linnartz} \label{section:0}
We would like to pay our gratitude and respects to our friend and colleague, Harold Linnartz, who passed away suddenly on Sunday, 31 December 2023 at the age of 58. Harold was a renowned figure in the field of molecular astrophysics, specifically excelling in molecular spectroscopy. His expertise extended to the application of spectroscopy in understanding the carriers of diffuse interstellar bands, exploring the spectroscopy and reactions of low-temperature solids relevant to interstellar ices, and delving into the photochemistry of PAHs, addressing challenges related to aromatic infrared bands in space.

Harold had envisioned continuing his groundbreaking research in these areas in the coming years, but it is not to be. We will miss him dearly as a colleague and, more importantly, as a friend.

%%% INTRODUCTION %%%
%%% INTRODUCTION %%%
\section{Introduction} \label{section:1}
Polycyclic aromatic hydrocarbons (PAHs) in the interstellar medium (ISM) are known for their ubiquity in a variety of astronomical environments including star- and planet-forming regions,\cite{Peeters2004, Bouwman2008, Calzetti2011, Cox2016} contributing up to $\sim$20\% of all cosmic carbon material in some objects.\cite{Tielens2008} They are widely accepted as carriers for a series of infrared (IR) features observed in the 3-20~\ce{\mu}m region known as the aromatic infrared bands (AIBs),\cite{Leger1984, Allamandola1989, Tielens2013} and are considered potential carriers for the diffuse interstellar bands (DIBs) present in the 4\,000-10\,000~{\AA} region which remain largely uncharacterised and have been a spectroscopic enigma for over a century.\cite{Herbig1995, Duley2006, Sarre2006, Cox2014} PAHs also play a major role in the chemistry and physics of photodissociation regions (PDRs), where far-ultraviolet (FUV, 6-13.6~eV) photons dominate, and serve as an important source of heat.\cite{Bregman1995, Pety2005, Joblin2011, Montillaud2013} Recent radio astronomy detections of indene (\textit{c}-\ce{C9H8})\cite{Burkhardt2021, Cernicharo2021} and 1- and 2-cyanonaphthalene (\textit{c}-\ce{C10H7CN})\cite{McGuire2021} in the Taurus Molecular Cloud-1 provide unambiguous evidence for the presence of PAHs in the ISM, but the chemical evolution of these interstellar PAHs, including their contribution to the carbon chemistry in astronomical environments, is still poorly understood.

When it comes to the formation of gas-phase astrochemically-relevant species, the bottom-up scenario is often invoked, where larger molecules are built from smaller species.\cite{Kaiser2021, McCabe2020, Bouwman2023} However, there is increasing evidence to suggest that the top-down scenario, \textit{i.e.}, the formation of small molecules from the destruction of larger species, is an important pathway, particularly in the case of PAHs.\cite{Berné2015, Zhen2018} Understanding the photodynamics of PAHs requires combined laboratory and theoretical studies that can quantify their photofragmentation pathways and their intrinsic dependence on photon wavelength and intensity. Multiphoton dissociation techniques are invaluable in experimentally investigating the chemistry and physics of cationic PAHs.\cite{Tielens2008, Oomens2003}

Recent experiments have revealed that photoprocessing of PAH cations leads to a number of (competing) fragmentation channels, including H-loss, \ce{H2}-loss, CH-loss, \ce{C2}-loss, and \ce{C2H2}-loss.\cite{Holm2011, Jochims1994, Simon2017, West2019, Panchagnula2020, Bouwman2016} Additionally, quantum chemistry studies have provided much insight into the reaction pathways of PAHs, emphasising the importance of hydrogen roaming and isomerisation in these fragmentation processes. \cite{Trinquier2017a, Trinquier2017b, Simon2017, Castellanos2018a} The extent of dissociation influences the resulting photodynamics and fragmentation pattern, with complete dehydrogenation of PAHs leading to species that comprise only carbon atoms organised in a ring or cage structure.\cite{West2014, Zhen2014a, Zhen2014b, Hrodmarsson2022} In the most recent of these works, Hrodmarsson \textit{et al.} investigated the photofragmentation patterns of three structural isomers of dibenzopyrene PAH cations (DBP, \ce{C24H14^.+}), each with a different molecular symmetry. Their results reveal that there are similarities in the laser-induced dissociation products of these three isomers despite the different geometric properties, with all three isomers breaking down to a series of carbon cluster fragments \ce{C_n} where \textit{n} = 11-15. However, their study did not discern fragments smaller than \ce{C11^.+} and thus did not investigate the formation routes of smaller PAH fragments.

In this work, we investigate the photodissociation of the small PAH cation coronene (\ce{C24H12^.+}) using time-of-flight mass spectrometry and quantum chemical computations to determine the structures and formation mechanisms of coronene photodissociation fragments down to 60~amu. We show that there are pathways to form carbon chains, clusters, and other relevant molecular fragments by irradiating PAHs, which have been already been observed in X-ray induced photodissociation.\cite{Huo2022, Garg2022, Lee2022} Lastly, we discuss the astrophysical implications of the fragmentation patterns mentioned here -- namely that they may provide clues about potential DIB carriers that come from the dissociation of the (partially) dehydrogenated parent.

%%% METHODS %%%
%%% METHODS %%%
\section{Methods} \label{section:2}
\subsection{Experimental} \label{section:2.1}
All mass spectra were obtained using our bespoke Instrument for the Photodynamics of PAHs (i-PoP). A detailed overview of the inner workings of the full system is provided by Zhen \textit{et al}.\cite{Zhen2014a} Here, only the relevant details are provided.

The i-PoP setup consists of a source chamber fitted with a Paul-type quadrupole ion-trap (QIT) that is connected \textit{via} a 2~mm skimmer to the detection chamber, which houses a reflectron time-of-flight (TOF) mass spectrometer. The base pressures of the chambers are ~$5.0\times10^{-8}$~mbar and ~$2.0\times10^{-9}$~mbar respectively. Commercially available coronene (\ce{C24H12}, Sigma Aldrich, 97\% purity) is sublimated in an oven in the source chamber at $\sim$393~K and ionised using electron ionisation with 70~eV electrons. The cations are then transported into the QIT \textit{via} an ion gate and trapped in a 1.25~MHz radiofrequency (RF) electric field (1300 and 2100~V\textsubscript{p-p}). A continuous flow of helium buffer gas is admitted directly into the ion trap to thermalise the ions and to reduce the size of the cloud.It has to be noted here that the ions formed by electron ionisation have a poorly defined internal energy distribution, but they cool down vibrationally and translationally by radiative cooling and collisions respectively with helium buffer gas before being excited by multiple laser pulses. The working pressures in the two chambers with the continuous helium flow are ~$1.6\times10^{-6}$~mbar (QIT) and ~$1.4\times10^{-8}$~mbar (TOF).

The ion cloud is then irradiated by multiple (up to 35) pulses of laser light from a tunable dye laser (LIOP-TEC, QuasAr2-VN), filled with the dye 4-(dicyanomethylene)-2-methyl-6-(4-dimethylaminostyryl)-4H-pyran (DCM) dissolved in ethanol, set to deliver 630~nm light. The dye was pumped by a Quanta-Ray Nd:YAG (INDI-40-10) laser operating at 532~nm and providing nanosecond long pulses at a repetition rate of 10~Hz. A wavelength of 630~nm is used to simulate vacuum-ultraviolet (VUV) photon induced dissociation processes in the ISM because its longer wavelength favours dissociation over multiple ionisation, allowing us to study the fragmentation patterns of coronene cations in greater detail. At shorter wavelengths, competition arises between the dissociation channels and multiple ionisation channels of the PAH cation.\cite{Zhen2016, Wenzel2020} Performing this experiment with UV photons would thus hinder or limit the study of continued fragmentation of the PAHs past the initial H-, \ce{H2}/2H-, \ce{CH}- or \ce{C2H2}-losses in a single experiment as a fraction of the remaining fragments eventually undergo double ionisation.\cite{Zhen2015} The coronene cation and its photofragments are excited using 630~nm photons. At the high local intensities of the nanosecond laser pulse ($\sim$0.5~MW/cm$^2$), non-linear absorption processes result in multiphoton excitation through virtual, rather than resonant, intermediate states.\cite{He2008} The ions can undergo intermolecular vibrational relaxation between (multi) photon absorptions, ensuring that double ionisation of fragments does not occur. This results in the high internal excitation required to initiate photo fragmentation processes in the trapped PAH cations.

A delay generator controls the timing sequence. Each cycle begins with a 2.7~s trap fill-time, followed by a 0.24~s stored waveform inverse Fourier transform (SWIFT) pulse, and 3.5~s of laser exposure. The SWIFT pulse is used to mass-selectively isolate ions. The total cycle time is set to 6.49~s, leaving 3.5~s seconds after the SWIFT pulse to irradiate the ions. An automated, mechanically-driven beam shutter determines the number of laser pulses interacting with the trapped ions. Following irradiation, the fragments are ejected into the TOF mass spectrometer and detected on a microchannel plate detector. The TOF mass spectrometer allows detection of the parent and daughter ions in a single mass spectrum. The data acquisition is controlled by LabVIEW programmes and data is analysed with a custom MatLab routine.\cite{MATLAB}

% IMAGE %
\begin{figure}[t]
\centering
  \includegraphics[width=\columnwidth]{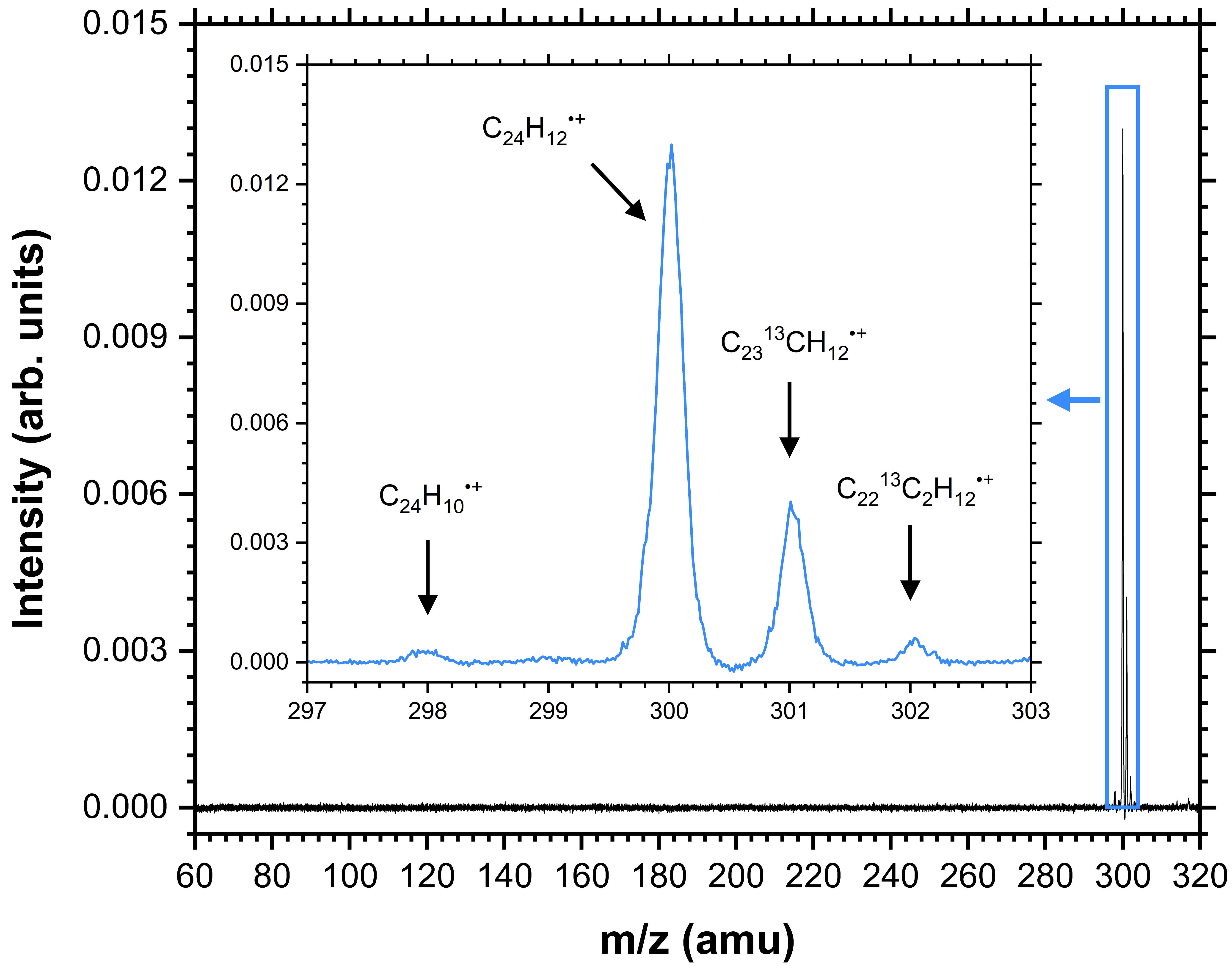}
  \caption{TOF mass spectrum of coronene radical cations trapped after sublimation, electron impact ionisation, and application of a SWIFT pulse. Inset: zoom-in around \textit{m/z} = 300 revealing the isotopic distribution and dehydrogenation of coronene cations, specifically including \ce{^13C}-containing species.}
  \label{Fig1}
\end{figure}

Two sets of experiments were conducted. First, we recorded mass spectra in the \textit{m/z} = 320-130 range (RF~=~2100~V, laser energy~=~1.1~mJ/pulse) to observe fragments in the high-mass range. We then recorded a second series of spectra in the \textit{m/z} = 320-80 range (RF~=~1300~V, laser energy~=~2.4~mJ/pulse) to observe fragments in the low-mass range (\textit{m/z} = 129-80).

% IMAGE %
\begin{figure*}[ht!]
\centering
  \includegraphics[width=\textwidth]{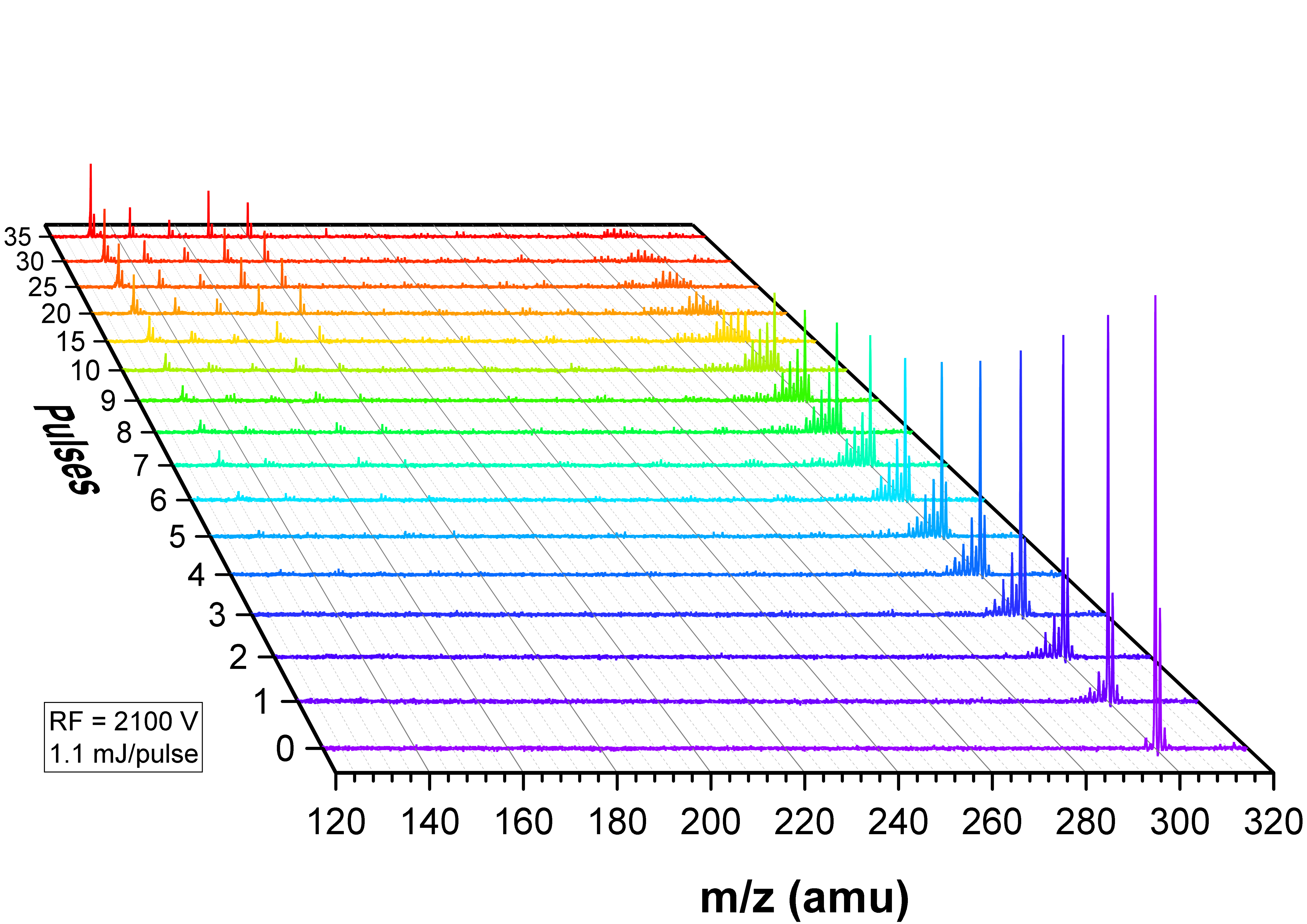}
  \caption{TOF mass spectra for coronene radical cations (\textit{m/z} = 300) irradiated from 0 to 35 laser pulses with a pulse energy of 1.1~mJ/pulse, showing spectra studying the high-mass (\textit{m/z} = 320-130) range.}
  \label{Fig2}
\end{figure*}

To account for the uncertainty of the absolute number of ions in the trap between the acquisition of one spectrum and the next, reference spectra (with no laser exposure) were recorded between each measurement. These reference spectra were used to normalise the peak areas of the fragmentation spectra (with laser exposure). When calculating the area of each peak, any \ce{^13C} contribution was corrected for and all areas were normalised to the parent ion signal (\textit{m/z} = 300). They were also subject to a detection threshold of 2$\sigma$ for the high-mass data set and 1$\sigma$ for the low-mass data. Peaks below these thresholds are indistinguishable from the noise level and are omitted from any discussion about overall trends in the fragmentation patterns.

\subsection{Computational} \label{section:2.2}
Investigation of the potential energy surface of the coronene cation and its fragments was performed with the Gaussian 16 \cite{Frisch2013} suite using density functional theory.\cite{Kohn1965} The M06-2X\cite{Grimme2011, Becke1993} functional and Pople's 6-311++G(3df,2pd) basis set were employed, based on the proven accuracy in estimating the difference in energy between isomers and between reaction barriers in hydrocarbons.\cite{Wiersma2021} The correct nature of transition states connecting reactants and products was checked using intrinsic reaction coordinate calculations. States with higher multiplicities were investigated in the case of partly dehydrogenated species.

%%% RESULTS AND DISCUSSION %%%
%%% RESULTS %%%
\section{Experimental results} \label{section:3}
In Fig.~\ref{Fig1}, a typical TOF mass spectrum is shown for coronene cations (\ce{C24H12^.+}, \textit{m/z} = 300) following a 2.7~s trap fill-time. The isotopologues and coronene fragments are labelled in the inset. To obtain this signal-to-noise ratio, mass spectra were recorded 70 times and co-added. The measured peaks in the range \textit{m/z} = 303-299 -- corresponding to isotopologues \ce{C22{}^13C2H12^.+} (302 amu, 5\%), \ce{C23{}^13CH12^.+} (301 amu, 26\%), and \ce{C24H12^.+} (300 amu) -- scale well with the natural \ce{^13C} abundance. As a result of electron impact ionisation, partially dehydrogenated coronene cations \ce{C24H11+} and \ce{C24H10^.+} (and their \ce{^13C} counterparts) are also present. Fig.~\ref{Fig1} confirms that the trapping time and efficiency were appropriate.

Accessing each fragmentation channel requires an intense nanosecond duration laser pulse which results in a non-linear response. This results in the intensities of individual peaks being highly sensitive to the laser energy, photon flux, and fluence. However, the general trends are robust and reproducible in experiments with different laser parameters.

Fig.~\ref{Fig2} shows the mass spectra for coronene radical cations irradiated with a laser pulse energy of 1.1~mJ/pulse in the \textit{m/z} = 320-130 (high-mass) range. In our experiments, laser irradiation of trapped ions leads to increased dissociation of the parent ion \textit{via} several competing channels. The first of these is the H-loss 
channel in which coronene loses hydrogen atoms to give fragments between \textit{m/z} = 299-288. With even more laser exposure, fragmentation channels between \textit{m/z} = 287-181 become accessible. In this range, CH losses become noticeable as laser exposure increases, and the CH-loss channel appears to be in competition with the H-loss channels. The 
first steps in the fragmentation of coronene cations following single-photon dissociative ionisation \textit{via} UV irradiation have been studied by Jochims \textit{et al.}\cite{Jochims1994, 
Jochims1999} They measure an appearance energy of 12.05 and 13.43~eV for the loss of H and \ce{H2} respectively. The \ce{H2} loss channel represents sequential loss of two 
hydrogen atoms.\cite{Castellanos2018a} For smaller PAHs, they also observe a \ce{C2H2}-loss channel which isn't accessible at the excitation energies employed in the Jochims \textit
{et al.} experiments (for coronene cations $\leq$13.7~eV).

While the dissociation channels of PAH cations commonly include the loss of \ce{C2} and \ce{C2H2}, \cite{Holm2011, Jochims1994, Simon2017, West2019} we observe a minimal presence of these channels. At higher laser fluences, a series of \ce{C_n^.+} cluster ions where \textit{n} = 10-15 appear, as well as smaller hydrocarbon fragments. The smallest fragment detected in these experiments is \ce{C7^.+}, in trace amount. We discuss each of these channels in turn, starting with the high-mass range results.

% IMAGE %
\begin{figure}[t]
\centering
  \includegraphics[width=\columnwidth]{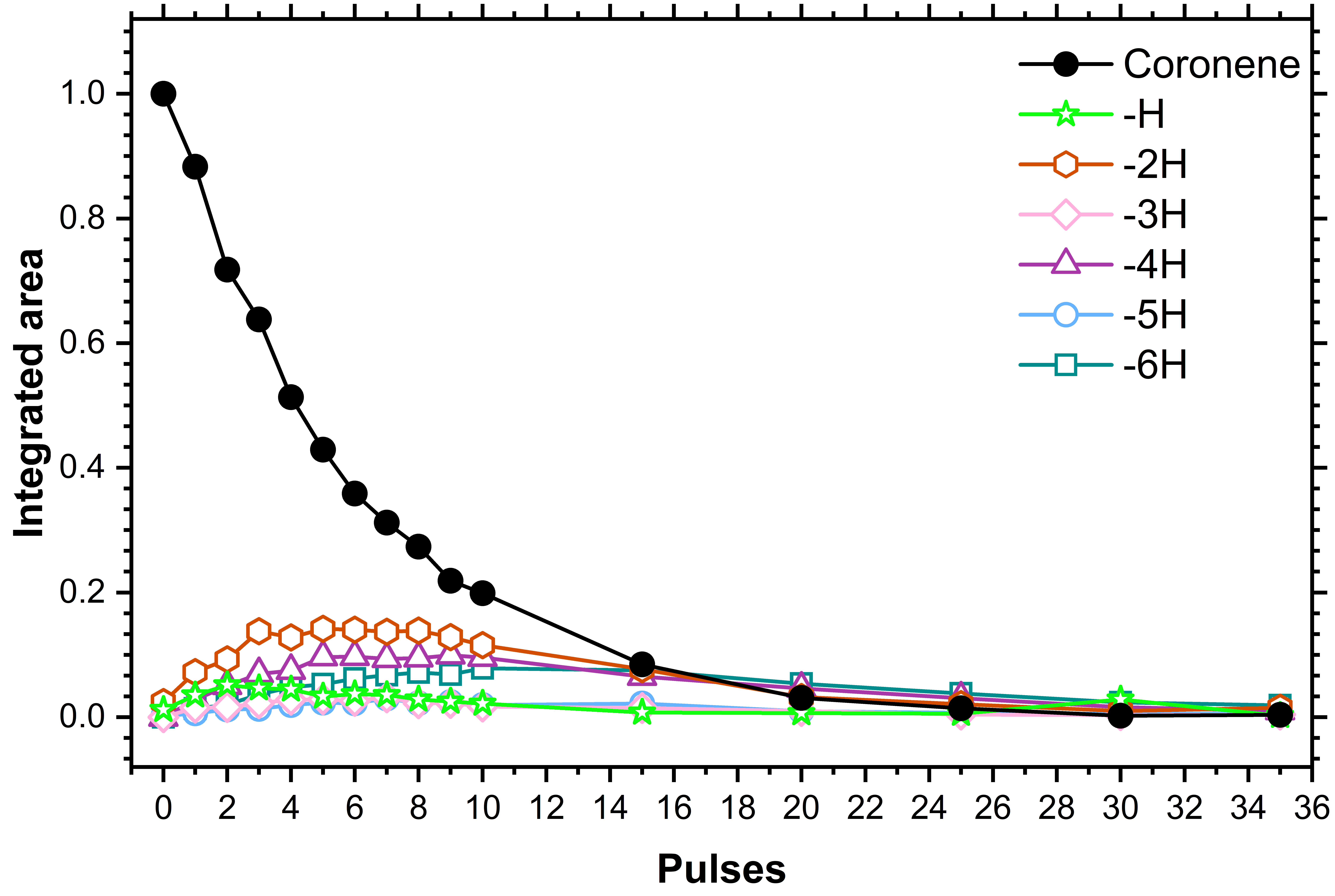}
  \caption{Integrated areas of select H-loss channels are shown with increased pulse exposure. The areas shown here are corrected for their \ce{^13C} contribution and normalised to the parent (\textit{m/z} = 300).}
  \label{Fig3}
\end{figure}

%%%%%%%%%%%%%%%%%%%%%%%%%%%%%%%%%%%%%%%%%%%%%%%%%%%%%%%%%%%%
\subsection{H-loss channel (\textit{m/z} = 300-288)} \label{section:3.1}
The relationship between the parent and fragment ions is clearly demonstrated in Fig.{~\ref{Fig3}}, which shows the relative integrated areas (corrected for \ce{^13C} contribution and normalised to the parent ion signal) of select H-loss channels (up to the loss of 6 hydrogen atoms). %Similar integrated areas for all dehydrogenation steps (up to the loss of 12 hydrogen atoms) are shown in the ESI.$^\dag$

\ce{C24H12^.+} (parent), \ce{C24H11+} (-H), and \ce{C24H10^.+} (-2H) are all present at 0 pulses as a result of electron impact ionisation. With increased irradiation, the area of the parent ion peak decreases while the H-loss peaks show an initial increase. The subsequent decrease of all H-loss signals to near-zero values is the result of further fragmentation of these species. In agreement with previous studies,\cite{Jochims1994, Joblin2004, Zhen2014a, Castellanos2018a} we conclude that the loss of an even number of hydrogen atoms is favoured over an odd number. There is a marked trend with the -4H channel becoming accessible after 1 laser pulse, the -6H channel at 2 pulses, the -8H channel at 3 pulses, the -10H channel at 4 pulses, and the -12H channel at 5 pulses. This demonstrates that the dehydrogenation process is sequential, with an increasing number of hydrogen atoms being individually stripped from the parent as the total energy absorbed by the parent increases. The dehydrogenation reaches completion when all twelve hydrogen atoms are removed, leaving only a bare carbon cluster (\ce{C24^.+}, \textit{m/z} = 288).

% IMAGE %
\begin{figure*}[t]
\centering
  \includegraphics[width=\textwidth]{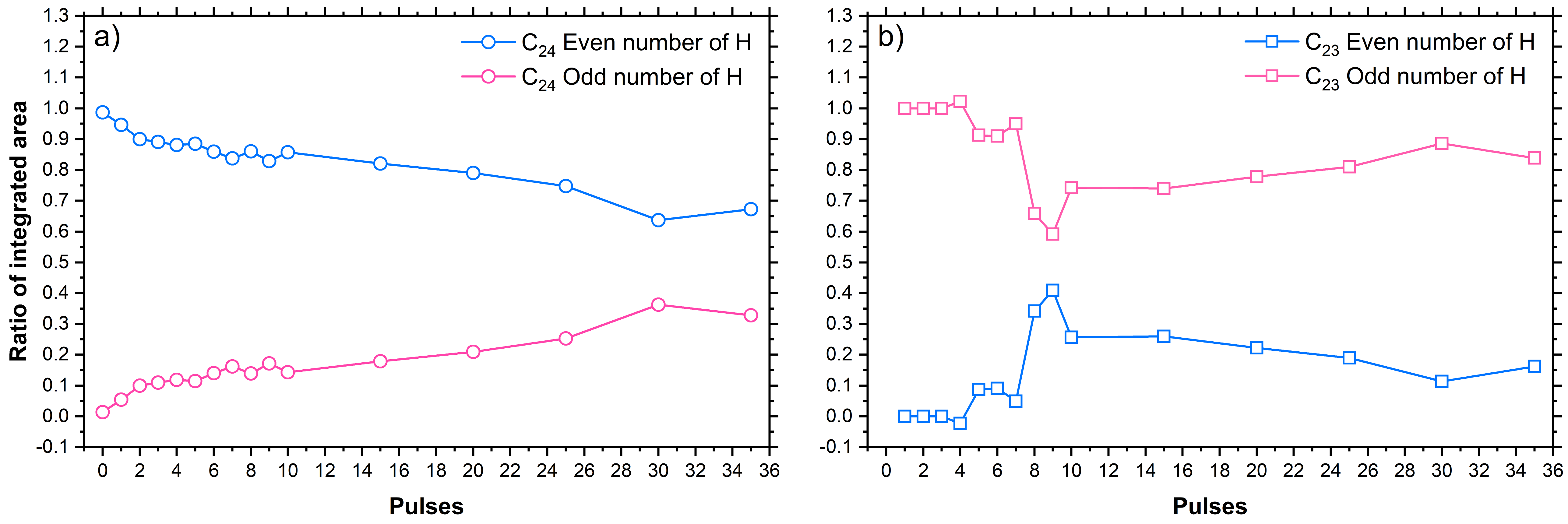}
  \caption{Ratio of integrated areas for even (blue) and odd (pink) species \ce{C_nH_x} where \textit{n} = (a) 24, and (b) 23. The areas shown here are corrected for their \ce{^13C} contribution and normalised to the parent (\textit{m/z} = 300).}
  \label{Fig4}
\end{figure*}

%%%%%%%%%%%%%%%%%%%%%%%%%%%%%%%%%%%%%%%%%%%%%%%%%%%%%%%%%%%%
\subsection{CH-loss channel (\textit{m/z} = 287-181)} \label{section:3.2}
The CH-loss channel is well-known in the photochemistry of small PAHs.\cite{Zhen2018, Lifshitz1997, West2014} At the start of the experiment, there is no evidence of a product at \textit{m/z} = 287, which would indicate the loss of a CH unit to give \ce{C23H11+}. We first see \ce{C23H11+} after 4 pulses of laser irradiation at 1.1~mJ/pulse. Since H-loss from the parent ion preferentially occurs in steps of -2H, there is a larger fraction of even masses in the range corresponding to ions with 24 carbon atoms (see Fig.~\ref{Fig4}a). The loss of CH from any of these species will result in a photoproduct with 23 carbon atoms and an uneven number of hydrogen atoms, as is reflected in Fig.~\ref{Fig4}b.

% IMAGE %
\begin{figure}[h]
\centering
  \includegraphics[width=\columnwidth]{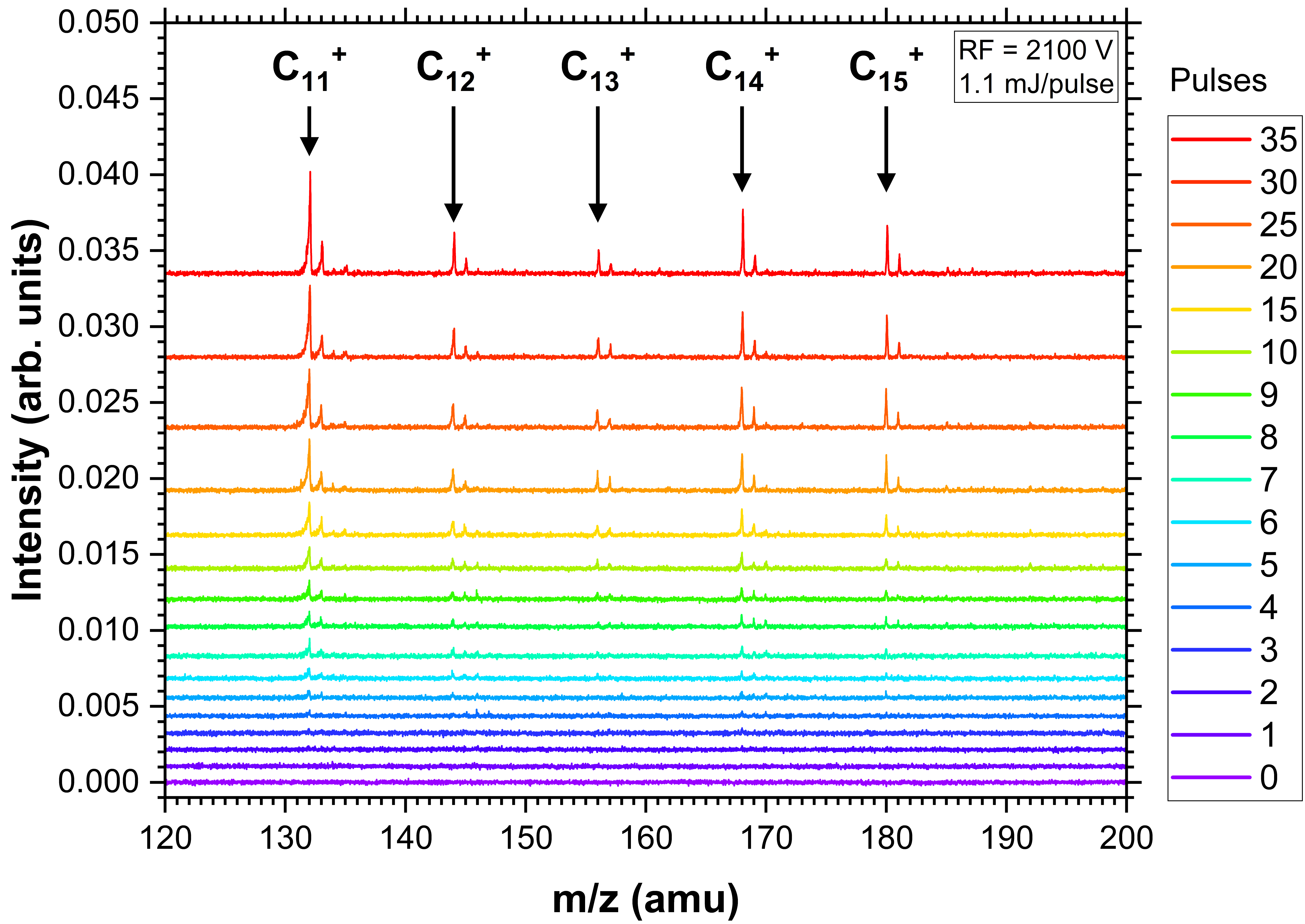}
  \caption{TOF mass spectra for coronene radical cations irradiated with 0 to 35 laser pulses at 1.1~mJ/pulse in the m/z = 320-130 (high-mass) range, showing the five carbon clusters of significance.}
  \label{Fig5}
\end{figure}

%%%%%%%%%%%%%%%%%%%%%%%%%%%%%%%%%%%%%%%%%%%%%%%%%%%%%%%%%%%%
\subsection{Carbon clusters (\textit{m/z} = 180-130)} \label{section:3.3}
Further down the mass spectrum, a series of high intensity peaks can be observed between \textit{m/z} = 180 and 130. These peaks correspond to small, bare carbon clusters that form from the parent species. The formation of such clusters has been seen in work by Zhen \textit{et al.} where they found that for clusters \ce{C_n^.+} where \textit{n}~$\geq$~32, the clusters were separated by 2 C atoms (\textit{i.e.}, \ce{C32^.+}, \ce{C34^.+}, \ce{C36^.+}...), and for \ce{C_n^.+} where \textit{n}~$\leq$~20, the cluster size differed by 1 C atom (\textit{i.e.}, \ce{C20^.+}, \ce{C19^.+}, \ce{C18^.+} ...).\cite{Zhen2014a} Fig.~\ref{Fig5} shows this is in agreement with our results, with strong signals for the following species separated by 1 C atom: \ce{C11^.+} (\textit{m/z} = 132), \ce{C12^.+} (\textit{m/z} = 144), \ce{C13^.+} (\textit{m/z} = 156), \ce{C14^.+} (\textit{m/z} = 168), and \ce{C15^.+} (\textit{m/z} = 180). Each of these \ce{C_n^.+} cluster species has an additional, small peak at one mass unit higher, \textit{i.e.} \ce{C_{n+1}^.+}, which may correspond to their singly-hydrogenated and \ce{^13C} counterparts. Hrodmarsson \textit{et al.} recently showed that, for carbon clusters up to \ce{C21+}, the contribution of \ce{C_nH+} decreases with increasing laser exposure, with only pure \ce{^13C} signals remaining. We expect similar behaviour in our experiment.\cite{Hrodmarsson2023}

\begin{figure}[h]
\centering
  \includegraphics[width=\columnwidth]{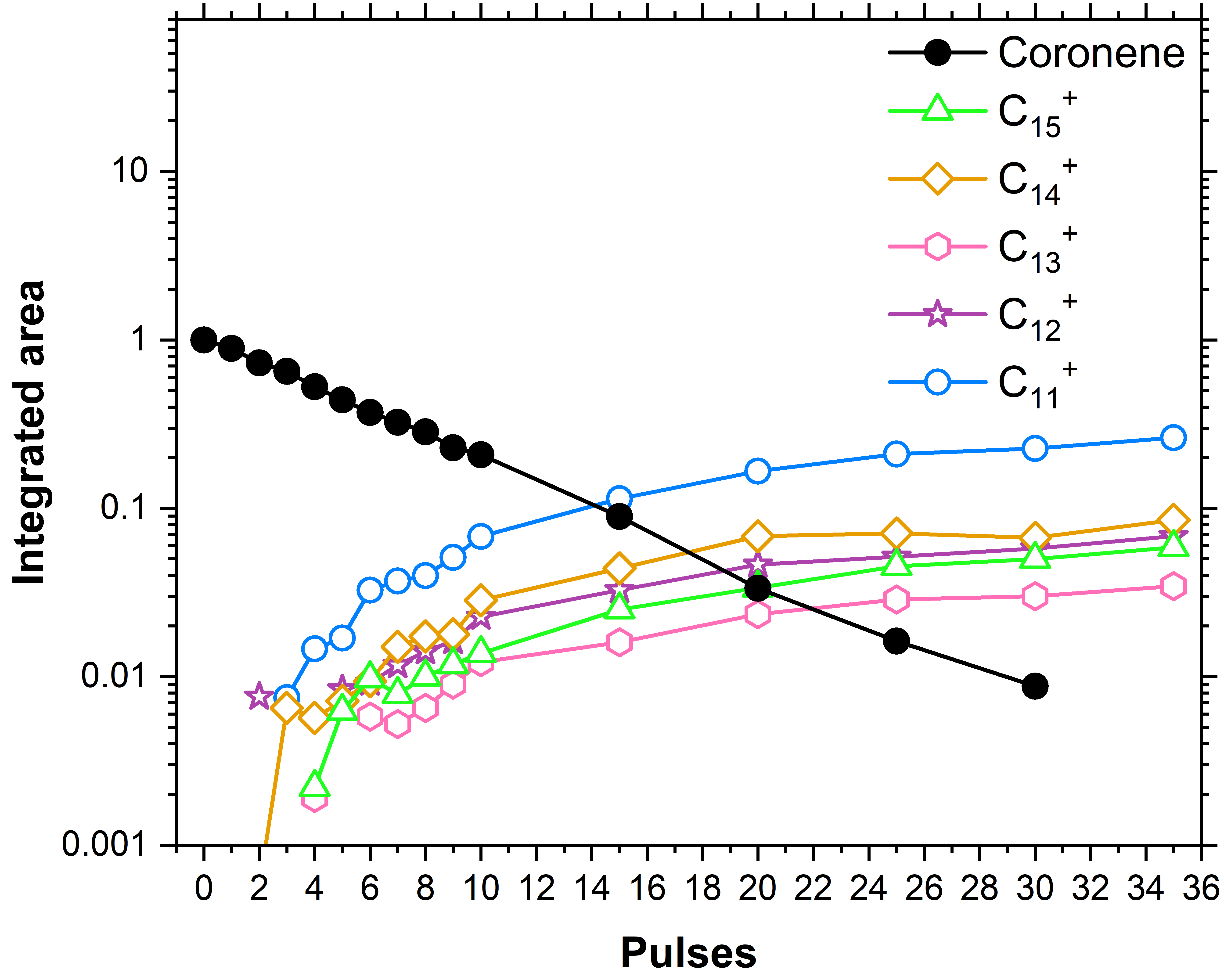}
  \caption{Integrated areas of the carbon cluster formation channels are shown with increased pulse exposure. The areas shown here are corrected for their \ce{^13C} contribution and normalised to the parent (\textit{m/z} = 300).}
  \label{Fig6}
\end{figure}

All the carbon clusters start to appear after partial dehydrogenation of the parent. \ce{C11^.+} appears immediately after 8 hydrogen atoms are lost from the parent (at 3 pulses). \ce{C12^.+}-\ce{C15^.+} are seen shortly after between 3-5 pulses, when full dehydrogenation takes place. These carbon clusters continue to grow over the course of the experiment at approximately the same rate (Fig.~\ref{Fig6}). Multi\-photon dissociation of smaller carbon clusters has been shown to preferentially lead to the loss of neutral \ce{C3}.\cite{vonHelden1993, Dynak2022} We note that, whilst the majority of the \ce{C11^.+} and \ce{C12^.+} signals we see here likely come directly from the dissociation of the coronene cation, there may be a minor contribution from the loss of a \ce{C3} unit from \ce{C14^.+} in the case of \ce{C11^.+}, and \ce{C15^.+} in the case of \ce{C12^.+}.

% IMAGE %

%%%%%%%%%%%%%%%%%%%%%%%%%%%%%%%%%%%%%%%%%%%%%%%%%%%%%%%%%%%%

% IMAGE %
\begin{figure*}[ht]
\centering
  \includegraphics[width=\textwidth]{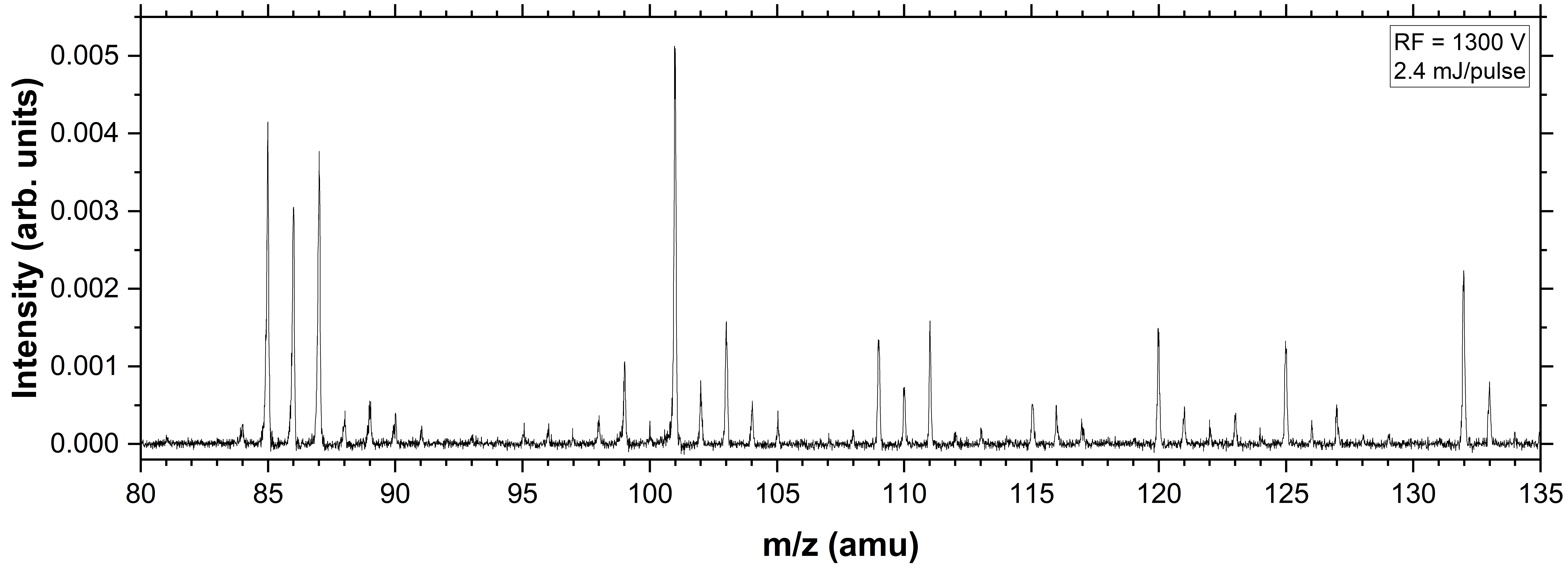}
  \caption{Zoom-in of the TOF mass spectrum of coronene radical cations irradiated with 20 pulses at 2.4~mJ/pulse in the \textit{m/z} = 135-80 range.}
  \label{Fig7}
\end{figure*}

%%%%%%%%%%%%%%%%%%%%%%%%%%%%%%%%%%%%%%%%%%%%%%%%%%%%%%%%%%%%
\subsection{Low-mass fragments (\textit{m/z} = 129-80)} \label{section:4.4}
To observe fragment species below 130~amu, a second set of experiments was performed where the laser pulse energy was set to 2.4~mJ/pulse and the amplitude of the RF signal of the ion trap was set to 1300~V. This data set gives insight into the behaviour of smaller dissociation fragments (\textit{m/z} = 129-80) while still enabling comparison with larger fragments of interest, \textit{e.g.}, \ce{C11^.+}. A representative spectrum is shown in Fig.~\ref{Fig7}. The full spectra for this data set are given in Fig. S1 and S2 in the ESI.$^\dag$

There are several prominent peaks of interest in this lower mass region: \textit{m/z} = 125, 120, 111, 109, 101, 87, 86, and 85. At 2 pulses of 2.4~mJ/pulse, \ce{C7H+}, \ce{C7H2^.+}, and \ce{C7H3+} all appear simultaneously. Only eight hydrogen atoms have been lost from the parent species at this stage, revealing that these low-mass fragments are some of the first to form following irradiation of coronene.

%The low-mass data set cannot be directly compared with the high-mass data set because, when comparing data sets collected at different RF values, the variations in ion trajectories and mass selection criteria introduce inconsistencies which can lead to differences in ion transmission efficiency, sensitivity, and selectivity between the two datasets. However,
It is possible to make qualitative comparisons between the two data sets. Two factors distinguish these low-mass fragments from the high-mass fragments. The first is the absence of bare carbon clusters at the low-mass region (\textit{i.e.}, \ce{C7^.+}, \ce{C8^.+}, and \ce{C9^.+}). The second is the difference in the hydrogenation of the each carbon backbone. \ce{C11H_{\textit{x}}+} to \ce{C15H_{\textit{x}}+} in the high-mass region only have one additional hydrogen atom (\textit{x}=1), whereas \ce{C7H_{\textit{x}}+}, \ce{C8H_{\textit{x}}+}, and \ce{C9H_{\textit{x}}+} have a series of hydrogen atoms (\textit{x}=1-9). Species with 10 carbon atoms exhibit chemistry that is common to both of these groups. Like the larger carbonaceous species, there is a dominant peak for the bare carbon cluster \ce{C10^.+}. However, unlike these larger species, there is a trail of hydrogenated \ce{C10H_{\textit{x}}} species reminiscent of the \ce{C7^.+} to \ce{C9^.+} group. The hydrogenation pattern is very different for each of the \ce{C7^.+} to \ce{C10^.+} fragments and hence, it is likely that these fragments do not feed into each other but rather originate from one (or more) species. The \ce{C11^.+} and \ce{C7H+} fragments continue to grow over the entire experiment, remaining the most dominant species after 35 pulses of laser exposure.

From earlier works\cite{Hrodmarsson2022, Hrodmarsson2023, Lee1997, vonHelden1993} and the data collected at an RF amplitude of 2100~V, similar behaviour is expected for the \ce{C_n^.+} species where n~$\geq~10$, \textit{i.e.}, following PAH dissociation, we mainly see the bare carbon cluster and its $^{13}$C and/or singly protonated counterpart (\ce{C_nH+}). This suggests that the appearance of a peak at \textit{m/z} = 125, which is comparable in intensity to the peak at \textit{m/z} = 120, is unusual. We propose that residual \ce{H2O} in our setup reacted with a lower mass fragment in the ion trap, resulting in ions which do not directly originate from parent (coronene) cation dissociation. For instance, the peak at \textit{m/z} = 125 may comprise \ce{C9H+} with an oxygen atom, \textit{i.e.}, \ce{C9HO+}. Similarly, the peak at \textit{m/z} = 127 is likely \ce{C9H3O+}. The assignment of low-mass peaks to specific species becomes challenging when there are unknown reactions occurring in the ion trap. It can be determined from atom deduction that \textit{m/z} = 111, 109, 87, 86, and 85 are likely \ce{C9H3+}, \ce{C9H+}, \ce{C7H3+}, \ce{C7H2+}, and \ce{C7H+} respectively. However, given the possibility of further reactivity in the ion trap, we cannot determine whether these ions originate directly from coronene cation dissociation, or form as a result of subsequent gas-phase reactions in the ion trap.

%%% RESULTS AND DISCUSSION %%%
{%%% DISCUSSION %%%
\section{Computational results and discussion} \label{section:4}
The experiments reveal a number of competing fragmentation channels including dehydrogenation through H- and \ce{H2}/2H-loss leading to \ce{C24H_x+} fragments, carbon skeleton fragmentation through CH-loss, and the production of carbon clusters (\ce{C_n^.+}, where \textit{n} = 11-15) and hydrocarbon species (\ce{C_nH_x+} where \textit{n} = 7-10). 
In this section, we provide a qualitative comparison and discussion of the experimental data with the results of density functional theory (DFT) calculations on the potential energy surface (PES) of coronene.

% IMAGE 8 %
\begin{figure*}
\begin{center}
  \includegraphics[width=\textwidth]{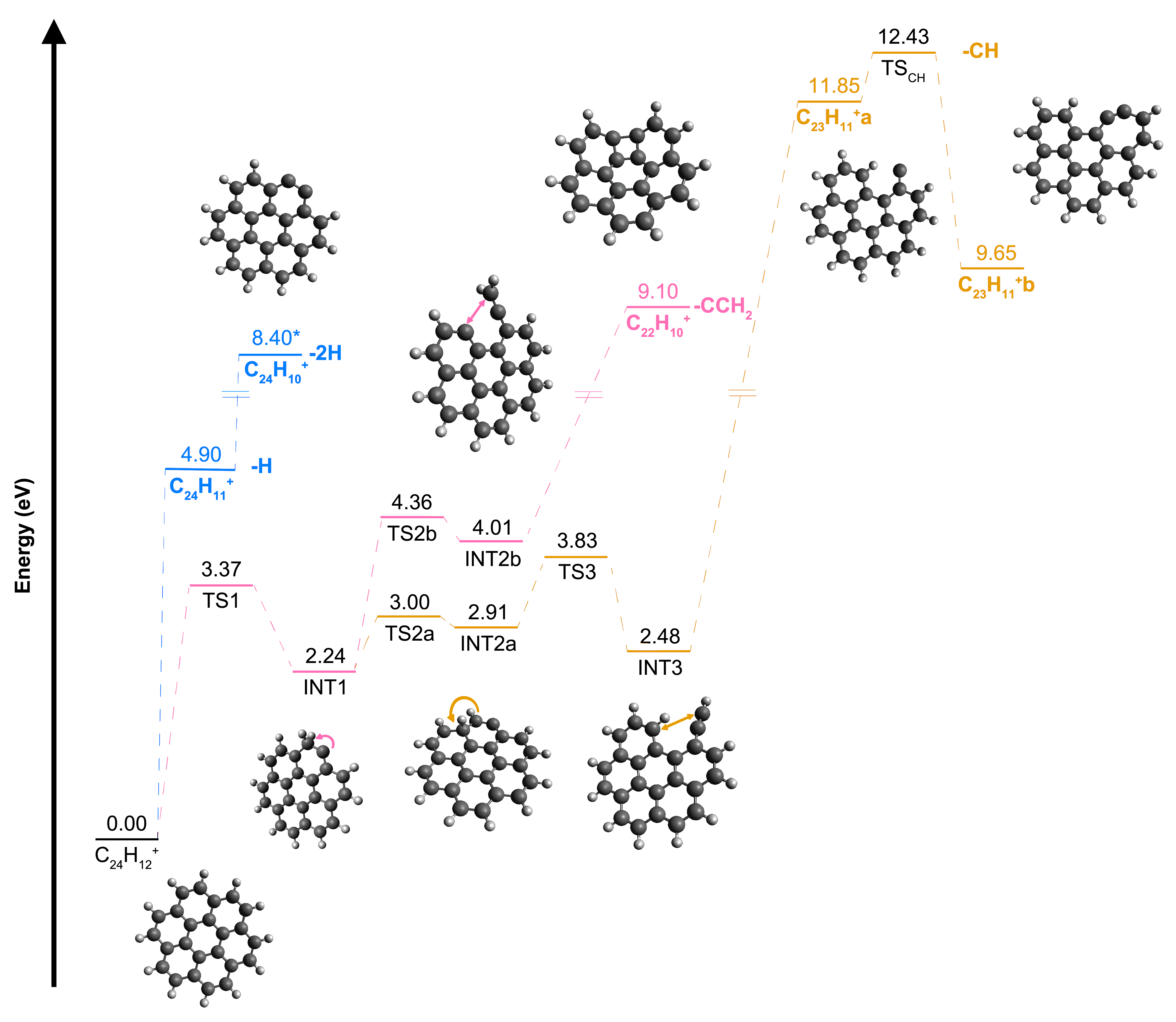}
  \caption{Potential energy surface illustrating several loss channels for the coronene cation: H- and 2H-loss (blue), \ce{CCH2}-loss (pink) and CH-loss (orange). The asterisk at \ce{C24H10+} is to signify that the BDE is calculated when the second H is removed from the partially dehydrogenated ring. If coming from another ring, the BDE is 1.5~eV higher\cite{Castellanos2018a}.}
  \label{Fig8}
\end{center}
\end{figure*}

%%%%%%%%%%%%%%%%%%%%%%%%%%%%%%%%%%%%%%%%%%%%%%%%%%%%%%%%%%%%
\subsection{H-, CH-, \ce{C2}-, and \ce{C2H2}-loss} \label{section:4.1}
Fig.~\ref{Fig8} shows the different pathways, energies of the products (relative to the coronene cation), and reaction barrier heights for H-loss, CH-loss, and \ce{C2H2}-loss. The H- and \ce{2H}-loss channels have already been computationally studied, but with a lower level of theory.\cite{Castellanos2018a, Trinquier2017a, Jolibois2005} The bond dissociation energies (BDE) for the direct loss of the first two hydrogen atoms are 4.90~eV and 8.40~eV respectively, in good agreement with the aforementioned studies. However, it is more likely that the first H-loss proceeds in two steps from \textbf{INT1} following hydrogen-roaming, with the same BDE. For comparison, the activation energy for H-loss from pyrene was measured by Lifshitz \textit{et al.} to be 4.6~eV, slightly lower than that calculated here for coronene.\cite{Lifshitz1997}

Subsequent isomerisation of \textbf{INT1} can lead to the formation of a vinylidene side group (\ce{-C=CH2}, \textbf{INT2b}, 4.01~eV). The BDE for \ce{CCH2}-loss was found to be 9.10~eV (5.09~eV from \textbf{INT2b}). The resulting co-fragment, \ce{C22H10+}, exhibits a non-planar structure induced by the presence of a four-membered ring. A similar structure was also observed in density-functional based tight-binding molecular dynamics simulations.\cite{Simon2017, Trinquier2017b} From \textbf{INT1}, two additional hydrogen-roaming steps lead to the formation of an ethynyl group (\ce{-C#CH}, \textbf{INT3}). CH-loss from this structure, with a BDE of 11.85~eV (9.37~eV from \textbf{INT3}), leads to the formation of \textbf{\ce{C23H11+}a}. A further, low-barrier isomerisation step (0.58~eV) leads to \textbf{\ce{C23H11+}b}, which contains a seven-membered ring that is lower in energy than \textbf{\ce{C23H11+}a} due to the formation of a triple bond between the two dehydrogenated carbon atoms in the ring.

From the energetics of the PES, it is evident that the H-loss channel is the most prominent at a low internal energy. This is confirmed by our experiments (Fig.~\ref{Fig2}) and other work.\cite{Zhen2016, Jochims1994} At higher internal energies, isomerisation-induced C-loss channels become accessible. The difference in our calculated BDEs suggests that \ce{CCH2}-loss should be more prominent than CH-loss. In the experiments presented in this work, we observe only a minor CH-loss channel. The competition between the CH- and \ce{CCH2}-loss channels studied here will be influenced by the possible presence of other isomerisation reactions (\textit{e.g.}, ring-opening and hydrogen-roaming) leading to different channels. Additionally, the photodissociation rate for the two channels can increase with different gradients as a function of the internal energy. A Monte Carlo simulation or transition state theory calculations of the photodissociation process may offer more insight into this issue, but are out of the scope of this paper.

%%%%%%%%%%%%%%%%%%%%%%%%%%%%%%%%%%%%%%%%%%%%%%%%%%%%%%%%%%%%
\subsection{Formation of carbon clusters (\textit{m/z} = 180-130)} \label{section:4.2}
Since \ce{C11^.+} is the first of the major fragment species to appear, it is most likely that \ce{C11^.+} forms directly from the parent or from an intermediary (partially-dehydrogenated) species (between \textit{m/z} = 299-181). Given that the \ce{C11^.+} peak continues to grow over the experiment as laser exposure increases but the peaks for \ce{C12^.+} to \ce{C15^.+} do not decrease in intensity, it is unlikely that \ce{C11^.+} forms from these larger carbon clusters. For this to happen, multiple high-energy dissociation steps are required. Since \ce{C11^.+} forms before the larger carbon clusters in the experiment, we can reasonably surmise that its formation route likely does not involve a \ce{C_n^.+} (where \textit{n} = 12-15) intermediate. All five carbon clusters appear to form directly from the dissociation of the parent coronene species.

To identify the potential formation pathways from the parent and to elucidate the structures of these carbon clusters, we used DFT. Two pathways were studied for the formation of the carbon clusters; one leading to the odd-numbered carbon rings (\ce{C11^.+}, \ce{C13^.+}, and \ce{C15^.+}) and the other to the even-numbered rings (\ce{C12^.+} and \ce{C14^.+}). Fig.~\ref{Fig9} presents the PES for these pathways, with \ce{C11^.+} being used to demonstrate the odd-numbered carbon cluster pathway (shown in pink) and \ce{C12^.+} being used to demonstrate the even-numbered carbon cluster pathway (shown in blue).

% IMAGE 9%
\begin{figure*}
  \includegraphics[width=\textwidth]{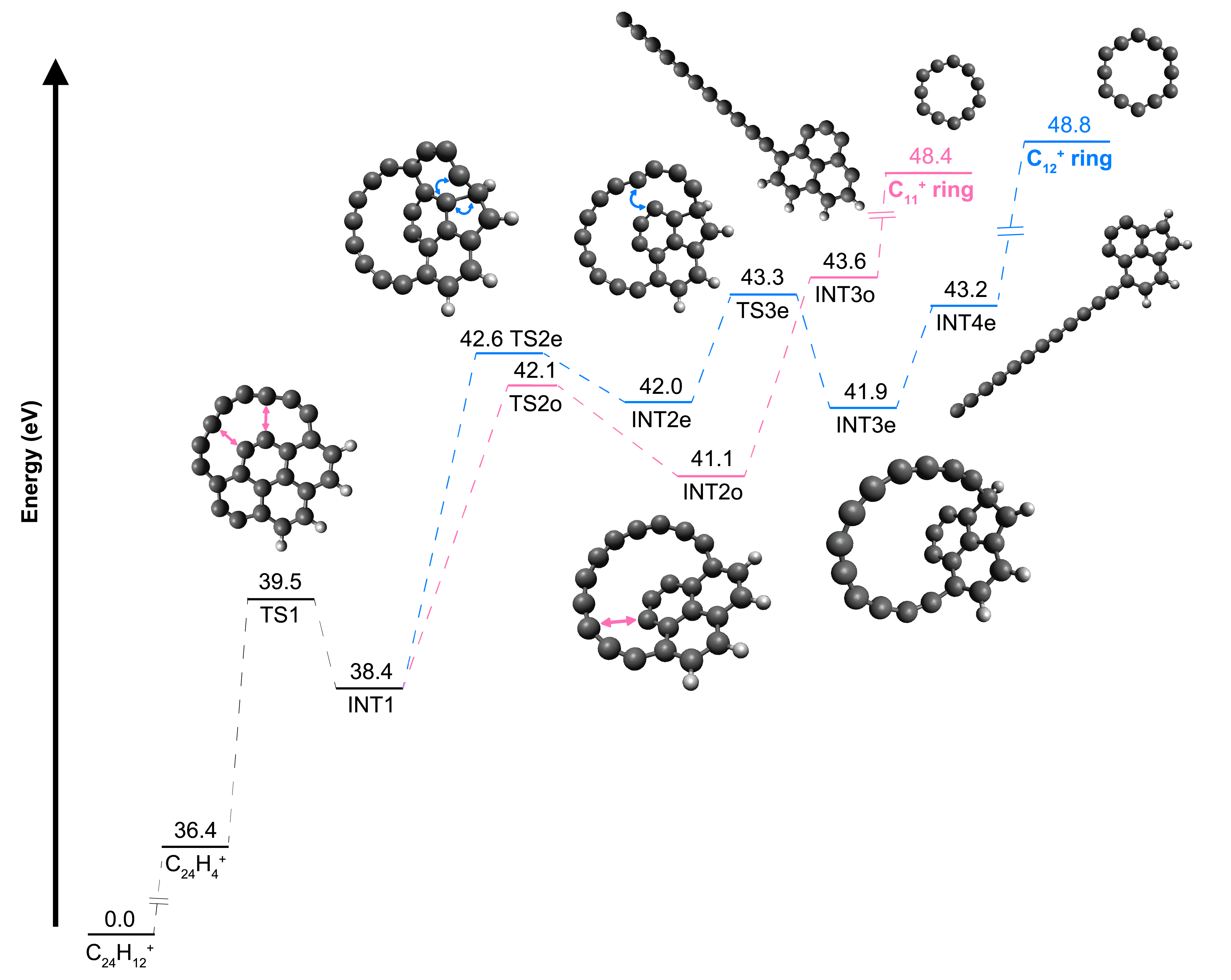}
  \caption{Potential energy surface showing the formation routes for \ce{C11^.+} (pink) and \ce{C12^.+} (blue). The corresponding structures for the transition states (TS) are shown, with atoms of interest highlighted (in pink for \ce{C11^.+} and blue for \ce{C12^.+}).}
  \label{Fig9}
\end{figure*}

Both pathways begin with the loss of 8 hydrogen atoms and share a common intermediate, \textbf{INT1}. This structure is the result of a reaction where two internal C-C bonds break open in a concerted manner (\textbf{TS1}), creating a chain of carbon atoms extending from a planar scaffolding as seen in other recent studies.\cite{Mackie2015, West2014} For \ce{C11^.+}, a third internal bond breaks open (\textbf{TS2o}) to give a species with a chain of 11 carbon atoms tethered to the central scaffolding on either end by a C=C bond. Breaking these tether bonds open is a barrier-less process and leads to the formation of \ce{C11^.+}. This species energetically prefers adopting a ring structure to a chain structure.\cite{vonHelden1993, Jones1999} The ring structure has a stabilising effect on the molecule and may explain why these \ce{C_n^.+} species do not seem to disintegrate even with increased laser exposure.

It is well-known that (partial or full) dehydrogenation of the parent initiates five-membered ring formation in the molecule, leading to structural irregularities and curling of the carbon skeleton.\cite{Berné2012, Bouwman2016, deHaas2017} This five-membered ring formation is invoked in the formation route for \ce{C12^.+} (and \ce{C14^.+}). Here, a 6-6~\textrightarrow~5-7 ring isomerisation step takes place (\textbf{TS2e}). In \textbf{TS3e}, the third internal bond to break is between a seven-membered ring and the central scaffolding (rather than a six-membered ring as in \ce{C11^.+}). Finally, the two tether bonds at each end of the chain break open to give \ce{C12^.+}, which also prefers to be a ring rather than a chain for the same reasons as \ce{C11^.+}.

Fig.~\ref{Fig10} illustrates how \ce{C13^.+}, \ce{C14^.+}, and \ce{C15^.+} can be formed from similar mechanisms as \ce{C11^.+} and \ce{C12^.+}. \textbf{INT3} (structure \textbf{1} in Fig.~\ref{Fig10}a) first loses its remaining four hydrogen atoms. The two six-membered rings connected to the carbon chain then isomerise to give two five-membered rings instead, creating a planar species with 13 carbon atoms extending in a tethered chain (structure \textbf{2}). This extension then breaks off to give a \ce{C13^.+} ring. For \ce{C15^.+}, one of the newly-formed bonds in the five-membered ring breaks, allowing the chain to grow by two further carbon atoms. Releasing this chain from the planar scaffolding gives a \ce{C15^.+} ring. The same mechanism of opening up the internal bond to the five-membered ring already present in \ce{C12^.+} leads to the formation of \ce{C14^.+}. We note that the \ce{C12^.+} formation route involves transition states that are at higher energies than comparable ones in the \ce{C11^.+} formation route. Hence, in a competition, \ce{C11^.+} formation will be favoured over \ce{C12^.+} formation at a typical internal energy. Additionally, \ce{C12^.+} and \ce{C13^.+} are antiaromatic in nature and require more energy to form. This may account for their relatively lower yields observed in our experiments (Fig.~\ref{Fig5}).

In recent work by Hrodmarsson \textit{et. al} on the fragmentation patterns of small PAHs, the dissociation of dibenzopyrene cations (\ce{C24H14^.+}) led to \ce{C_n^.+} (where \textit{n} = 11-15) formation early in the experiment, with additional peaks for hydrogenated \ce{C_n^.+} (\ce{C_nH_x+}, where \textit{x} = 1-2) that disappeared as the amount of energy deposited into the system increased. In the dissociation of dicoronylene cations (\ce{C48H20^.+}), \ce{C11^.+} was detected as a key fragment immediately after full dehydrogenation of the dicoronylene parent, similar to what we observe in the present work. If we also consider the results of our experiments on coronene, a compact PAH, we can suggest that compact PAHs (such as coronene and dicoronylene) require dehydrogenation before the formation of carbon clusters, while for irregular or catacondensed species (such as dibenzopyrene and iso\-violanthrene) there is greater competition between H-, CH-, and \ce{C2H2}-loss, and \ce{C_n^.+} formation.

% IMAGE 10 %
\begin{figure}
  \includegraphics[width=\columnwidth]{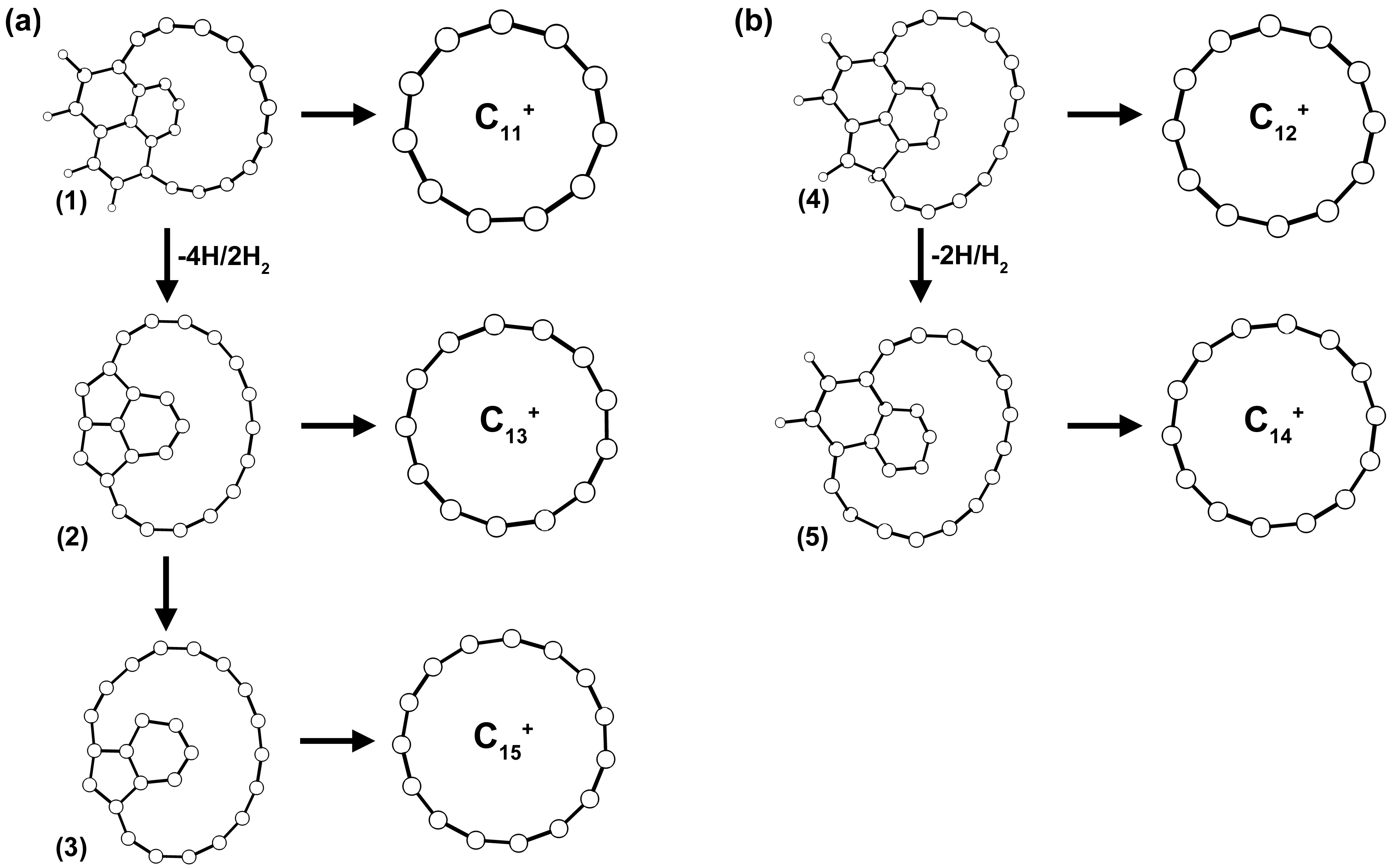}
  \caption{Fragmentation map with structures illustrating formation routes for (a) odd-numbered carbon clusters \ce{C11^.+}, \ce{C13^.+}, and \ce{C15^.+}, and (b) even-numbered carbon clusters \ce{C12^.+} and \ce{C14^.+}.}
  \label{Fig10}
\end{figure}

%%%%%%%%%%%%%%%%%%%%%%%%%%%%%%%%%%%%%%%%%%%%%%%%%%%%%%%%%%%%
\subsection {Low-mass fragments} \label{section:4.3}
In the low-mass range, the observed carbonaceous species are hydrogenated and do not appear to fragment further, which is indicative of a \ce{C_nH_x+} chain (where \textit{n} = 7-10) with one or more terminal hydrogen atoms. In the present work, we consider that \textit{m/z} = 87, 86, and 85 are \ce{C7H3+}, \ce{C7H2+}, and \ce{C7H+} respectively (Fig.~\ref{Fig7}). Here, we propose fragmentation pathways to each of these species from the coronene cation.

The first step in the formation of \ce{C7H+} is the loss of 8 hydrogen atoms from the parent, the same as when forming the carbon clusters \ce{C_n^.+} where \textit{n} = 11-15 (Fig.~\ref{Fig11}). A hydrogen atom then roams two carbons away onto the adjacent ring (\textbf{TS1} and \textbf{TS2}). A 6-6~\textrightarrow~5-7 ring isomerisation then takes place (\textbf{TS3}), following which two internal bonds open (\textbf{TS4} and \textbf{TS5}) to give a chain of seven carbon atoms and one hydrogen atom extending from a planar scaffolding, as seen in the formation of the carbon clusters \ce{C_n^.+}. This chain then frees itself from the scaffolding (\textbf{TS6}) to give \ce{C7H+} (47.5~eV), a linear chain with one terminal hydrogen on an \textit{sp}-hybridised carbon. The initial transition states involved in the \ce{C7H_x+} formation route are at similar energies to those in the \ce{C11^.+} route, and it may appear as though the ring- and chain-formation channels are in competition. However, due to the extent of dehydrogenation, some of the transition states and intermediates leading to \ce{C7H+} have very similar energies when higher spin multiplicities (s = 4, 6, and 8) are considered. These species have possible multi-reference character and single-determinant methods such as DFT are not well suited to elucidate these reaction paths in a quantitative manner. We also cannot exclude the possibility of additional isomerisation reactions occurring. However, we include our calculations here in an effort to offer a qualitative overview of the processes involved in key fragments of the low-mass region.

% IMAGE 11 %
\begin{figure*}
  \includegraphics[width=\textwidth]{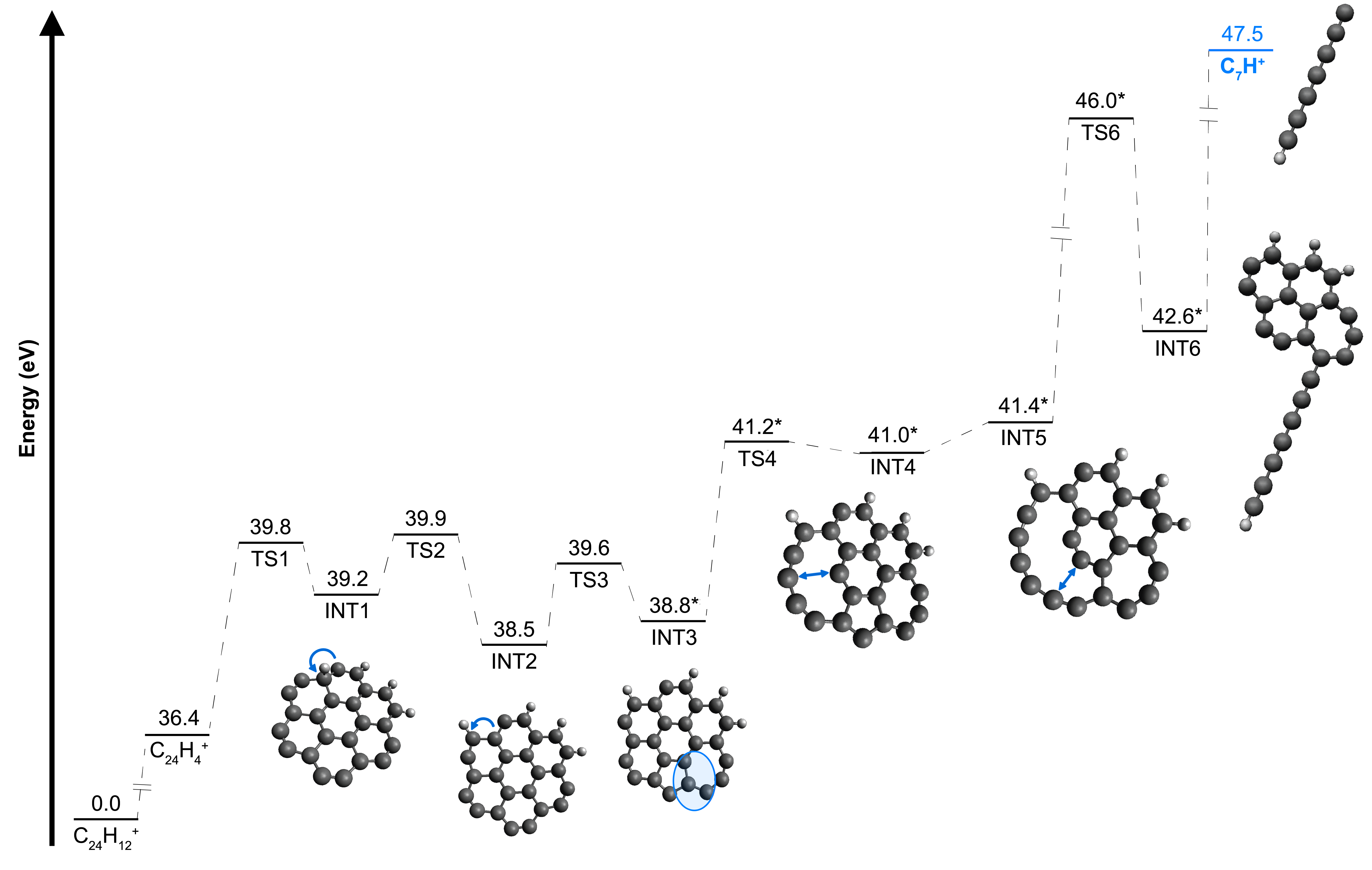}
  \caption{Potential energy surface showing a potential formation route for \ce{C7H+}. The blue arrows on the intermediate structures (\textit{e.g.}, INT4) show the changes in the transition state (TS) leading to the specific intermediate (\textit{e.g.}, TS4). Asterisks indicate structures where optimisation to different spin states (doublet, quadruplet, and sextuplet) give rise to very similar total energies, within 0.2~eV, highlighting their multi-reference character.}
  \label{Fig11}
\end{figure*}

Fig.~\ref{Fig12} illustrates how \ce{C7H2+} and \ce{C7H3+} can be formed from a similar mechanism. Before the \ce{C7H+} chain breaks from \textbf{INT6} (structure \textbf{1} in Fig.~\ref{Fig12}), a hydrogen atom migrates onto the chain (structure \textbf{2} in Fig.~\ref{Fig12}). This then dissociates from the planar scaffolding to give \ce{C7H2+}, a linear chain with two hydrogen atoms at one end on an \textit{sp$^2$}-hybridised carbon. If another hydrogen migration step takes place before the chain breaks, \ce{C7H3+} may be formed. This species also has a linear structure with two terminal hydrogen atoms at one end, and another hydrogen atom at the other end.

% IMAGE 12 %
\begin{figure}
  \includegraphics[width=\columnwidth]{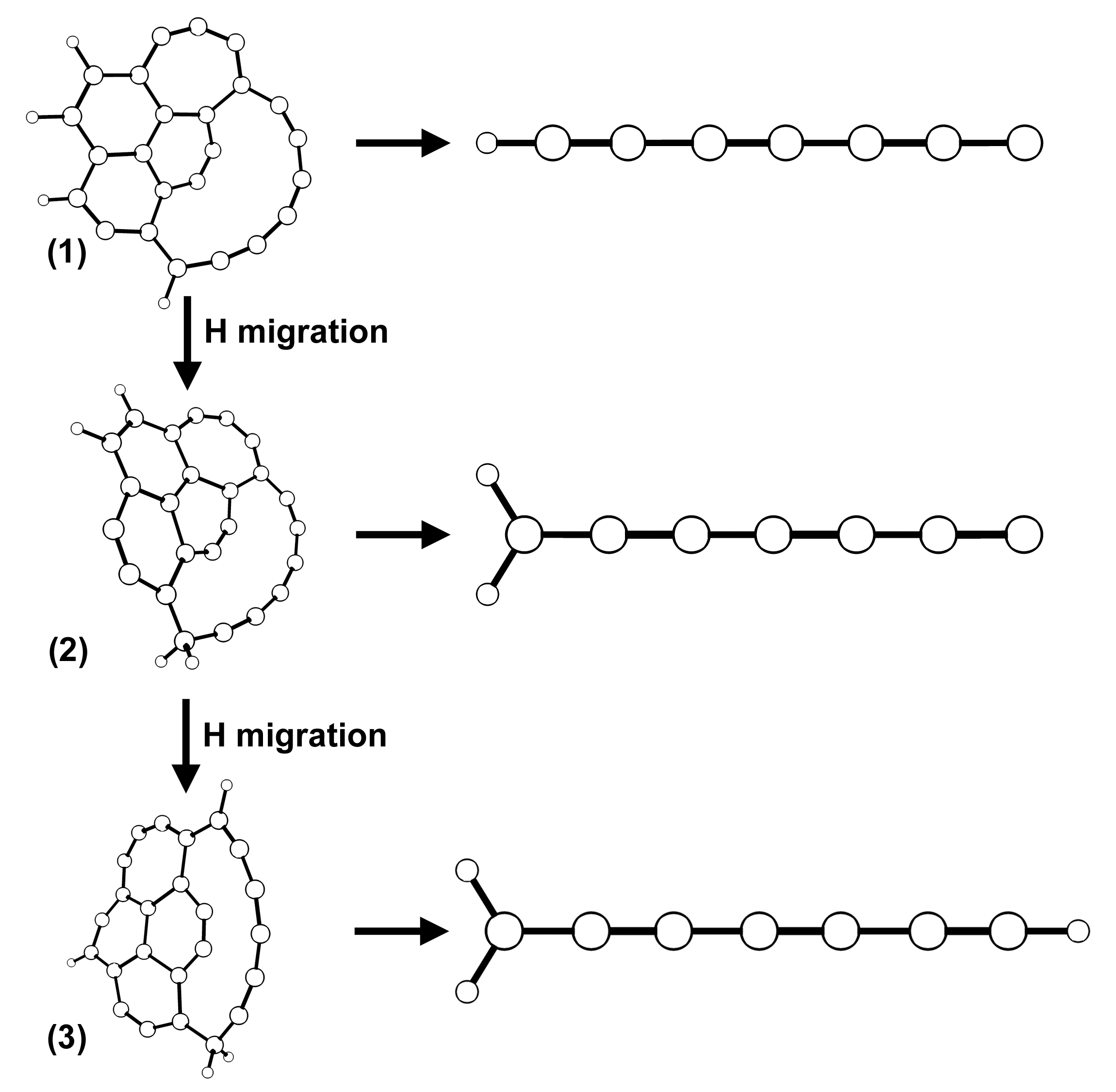}
  \caption{Fragmentation map with structures illustrating formation routes for \ce{C7H+}, \ce{C7H2^.+}, and \ce{C7H3+}.}
  \label{Fig12}
\end{figure}}

%%% ASTROPHYSICAL IMPLICATIONS %%%
{%%% ASTROPHYSICAL IMPLICATIONS %%%
\section{Astrophysical implications} \label{section:5}
The IR spectra of many regions in space -- including carbon-rich planetary nebulae, photodissociation regions (PDRs) associated with ionised gas regions powered by O-type stars and reflection nebulae illuminated by late B-type stars, planet-forming disks associated with young stars, starburst regions associated with galactic nuclei, and the general ISM of galaxies -- show broad emission features commonly ascribed to fluorescence by ultraviolet-pumped PAH molecules.\cite{Tielens2008} These PAHs are an important component of the ISM of galaxies, locking up $\sim$20\% of the elemental carbon and affecting the energy balance and ionisation equilibrium of atomic and molecular clouds.\cite{Tielens2008} In addition, PAHs may also be carriers of some DIBs, a set of $\sim$500 visual absorption features in stellar spectra arising from molecular electronic transitions.\cite{Herbig1995, Snow2014}

Besides PAHs, the Buckminster fullerene (\ce{C60}) and its cation have been detected in space through their vibrational IR emission spectra and electronic absorption spectra.\cite{Cami2010, Campbell2015, Berné2013, Spieler2017, Linnartz2020} IR spectroscopic observations have revealed that the abundance of \ce{C60} increases close to strong sources of ultraviolet (UV) radiation, while the abundance of PAHs decreases.\cite{Berné2012} This likely reflects the importance of top-down chemistry where UV photolysis of large PAHs leads to H- and \ce{C2}-loss, as well as isomerisation of the resulting carbon structures to \ce{C60}.

Understanding the origin and evolution of interstellar PAHs and their relationship to other organic compounds in space is therefore a key objective in the field of astrochemistry. Photolysis of PAHs by UV photons is of paramount importance as this will limit the PAH lifetime under interstellar conditions, and hence their putative contribution to the organic inventory of newly forming planets.\cite{Micelotta2011} UV processing may also serve as a source of small hydrocarbon radicals in UV-rich environments that can then be broken down further to smaller species, which in turn act as "feedstock" for the formation of larger species.\cite{Pety2005} Hence, there is a strong need to study PAH photodissociation processes and identify their fragmentation pathways. This work expands on the existing knowledge of coronene and other small PAH fragmentation and brings new insights into the formation pathways of astronomically relevant (hydro)carbon chains and rings.

Early studies on the photochemistry of PAHs in the ISM concentrated on identifying the first steps in PAH fragmentation, involving competing channels of H-, \mbox{\ce{H2}/2H-,} and \ce{C2H2}-loss.\cite{Jochims1994, Jochims1999} More recent imaging photoelectron photo\-ion coincidence spectroscopy experiments have determined the activation energies and changes in entropies for these channels by modelling the experimental breakdown diagrams.\cite{West2018} The present work adds to this by identifying an additional first-step fragmentation channel: the loss of CH from the coronene parent. As discussed in Section~\ref{section:4.2}, we link this to the formation of pentagons. Spectroscopic studies have demonstrated pentagon formation for two- and three-ring PAHs.\cite{Bouwman2016, Petrignani2016, deHaas2017, Banhatti2022} In our multi\-photon study on coronene dissociation, two additional CH-loss steps lead to a sumanene-like species with a bowl-shaped molecular structure. This CH fragmentation pathway unlocks a bottom-up chemical route toward the formation of \ce{C60} in the ISM.

Multiphoton processes in the ISM play a crucial role in understanding IR emissions and dissociations from large interstellar PAHs, typically consisting of 40-50 carbon atoms.\cite{Croiset2016} These processes become particularly relevant in the vicinity of hot stars found in star- and planet-forming regions.\cite{Andrews2016, Lange2021} In these experiments, various pulses of radiation occur, with each pulse typically involving absorption of a few low-energy photons by the PAH cations and their fragments. This allows ample time for relaxation and dissociation to occur \textit{via} low-energy channels. Therefore, the explored fragmentation mechanism is undeniably pertinent to astrophysics. Moreover, it is worth noting that PAHs are detected in diverse astrophysical settings where they are subjected to radiation fields with varying spectral energy distributions. For instance, some radiation fields exhibit higher UV fluxes compared to others. Specifically, in the vicinity of massive stars within PDRs, a coronene molecule could absorb a photon every ~10\,000~s.\cite{Tielens1985} However, within the lifespan of a region spanning 100\,000~yr, the cumulative occurrences of photon absorption events amount to 100\,000\,000, rendering even seemingly improbable events likely.\cite{Andrews2016, Montillaud2013} This underscores the importance of considering the broader context of radiation environments when studying PAH molecules in astrophysical contexts.

As mentioned above, photofragmentation experiments of large PAHs start with H-loss. The bare carbon clusters then fragment through the loss of \ce{C2} units.\cite{Berné2012, Zhen2014b} These experiments reveal that the carbon clusters display the same "magic numbers" as photofragmentation experiments on the fullerenes, \ce{C60} and \ce{C70}, supporting the isomerisation of these carbon clusters to \ce{C60} in a top-down route towards fullerenes. The bottom-up chemical route to \ce{C60} initiated by CH-loss from coronene, identified in this study, provides an alternative to this top-down pathway where "bowl" formation would be followed either by ion-molecule reactions with \ce{C2H2} -- akin to the routes towards naphthalene in dense molecular cloud cores \cite{McGuire2021} -- or through formation of PAH clusters followed by isomerisation.

The main objective of this study is to follow photolysis processes down to the smallest cationic fragments. In our experiments, we identify the facile formation of carbon species in the range of 7-10 and 11-15 carbon atoms. Based on our DFT studies, we link these to ring and chain species. The DFT results identify isomerisation of two adjacent hexagons to a pentagon-heptagon structure in the coronene species as a key intermediary step in their formation. Such isomerisation is a well-known process in naphthalene--azulene conversion,\cite{Bouwman2016, Dyakov2006} but its influence on the breakdown of larger PAHs has not been previously recognised in astrophysical studies. While the \ce{C11^.+}-\ce{C15^.+} linear chains do not have a permanent dipole moment, some of their hydrides as well as their non-linear isomers do and could be detected through their rotational fingerprint in the sub-millimetre regime. In this respect, rotational emission spectra have revealed the presence of linear carbon chains as large as \ce{C8H} in interstellar molecular clouds.\cite{Suzuki1986, Bell1999} The vibrational spectra of small \ce{C_n} chains (where \textit{n} $\leq$ 5) have been discovered in the outflow from the late-type giant, CW Leonis (IRC +10216),\cite{Bernath1989} and the launch of the James Webb Space Telescope may provide the best opportunity to detect these species directly in the ISM.

Generally, interstellar carbon chains are thought to result from ion-molecule reactions involving acetylene (\ce{HC#CH} and its derivatives.\cite{Chabot2013} However, breakdown of PAHs, such as coronene, may provide an alternative route toward these species, particularly in photodissociation regions and the diffuse ISM where build-up of larger carbon chains from small hydrocarbon radicals may be inhibited by the strong UV field. The relative proportions of the small carbon clusters and hydrocarbon species in our experimental study may be affected by the opening up of efficient, allowed, electronic radiative cooling processes that compete well with fragmentation due to the large Einstein A coefficients involved.\cite{Iida2022} These same electronic cooling processes would also promote the survival of these hydrocarbon chains in the UV-rich ISM of galaxies, greatly increasing their abundance. We note that, if these electronic transitions connect to the ground state, these species would be promising candidates for the carriers of the DIBs. While earlier studies on such species have not been successful in identifying carriers,\cite{Rice2013, Linnartz2010} photochemical studies as reported here may serve as a useful guide toward potential DIB candidates.}

%%% CONCLUSIONS %%%
%%% CONCLUSIONS %%%
\section{Conclusions} \label{section:6}
Mass spectra of the coronene cation (\ce{C24H12^.+}) were recorded using time-of-flight mass spectrometry. A series of spectra with increasing laser exposure were collected between \textit{m/z} = 300-60 at two different laser pulse energies, 2.4~mJ/pulse and 1.1~mJ/pulse. The spectra showed key photodissociation fragments in the high-mass region (\textit{m/z} = 300-130) and the low-mass region (\textit{m/z} = 129-60).

The high-mass region was dominated by the bare carbon clusters, \ce{C_n^.+} where \textit{n} = 11-15. DFT calculations indicate that these species are likely formed through a series of internal C-C bond rearrangements and dissociations. The same is true for the low-mass region, which showed a number of smaller hydrocarbon chains. \ce{C7H+} was the most abundant species, and its formation pathway involves hydrogen-roaming. Pentagon formation through isomerisation reactions involving a 6-6~\textrightarrow~5-7 transformation plays a key role in the formation of the hydrocarbon rings and chains observed in this study. %Pentagon formation is an important step in the formation of both the bare carbon clusters and the hydrocarbon chains observed in this study.

This work presents an alternative to the well-studied top-down formation route for \ce{C60}, and expands on existing knowledge of coronene and other small, astronomically relevant PAHs. This work may benefit from more complex computational studies to better determine the energies, structures, and pathways to the low-mass fragments presented here.

%%% CONFLICTS OF INTEREST %%%
\section*{Conflicts of interest}
There are no conflicts to declare.

%%% ACKNOWLEDGMENTS %%%
\section*{Acknowledgements}
S.P. acknowledges the European Union and Horizon 2020 funding awarded under the Marie Sk{\l}odowska-Curie action to the EUROPAH consortium (grant number 722346). H.R.H. acknowledges funding from the European Union and Horizon 2020 research and innovation programme (grant number 838372).
H.L. acknowledges the Netherlands Research School for Astronomy (Nederlandse Onderzoekschool Voor Astronomie, NOVA) and the Netherlands Organisation for Scientific Research (Nederlandse Organisatie voor Wetenschappelijk Onderzoek, NWO). J.B. acknowledges the NWO for a VIDI (grant number 723.016.006). A.G.G.M.T. acknowledges the NWO for a Spinoza Prize (Spinozapremie), and the NWO Exact and Natural Sciences (Domein Exacte en Natuurwetenschappen, ENW) for the use of supercomputer facilities (grant numbers 16638 and 17676).  This work was supported in part by NASA's Solar System Exploration Research Virtual Institute (SSERVI): Institute for Modeling Plasma, Atmosphere, and Cosmic Dust (IMPACT) This article is based upon work from European Cooperation in Science and Technology (COST) Action NanoSpace, CA21126, supported by COST.

\balance

\renewcommand\refname{References}

%%% REFERENCES %%%
\bibliography{bibliography.bib}
\bibliographystyle{bibliography.bst}

\end{document}